# Laser fabrication of vibrating microcavities on silicon: Formation dynamics and frequency tunability

Belen Valiente[1,*] Gonzalo Gomez-Munoz[1], Mario Garcia-Lechuga[1], Irene Solana[1], Fernando Agulló-Rueda[2], David Grojo[3], Angel Rodriguez-Fernandez[4], Javier Solis[1], and Jan Siegel[1,*]

[1] Laser Processing Group, Instituto de Optica (IO), CSIC, Serrano 121, 28006 Madrid, Spain
[2] Instituto de Ciencia de Materiales de Madrid (ICMM), CSIC, 28049, Madrid, Spain
[3] Aix-Marseille Université, CNRS, LP3 UMR 7341, Marseille 13009, France
[4] Diamond Light Source, Harwell Science & Innovation Campus, Didcot, OX11 0DE, UK

[*]Corresponding author emails: belen.valiente@io.cfmac.csic.es, j.siegel@csic.es

## Abstract:

Laser processing of silicon has long since become a standard method for wafer dicing, surface texturing, re-crystallization after ion implantation, as well as surface amorphization, to name a few examples. Here, we introduce a novel concept, the fabrication of vibrating microcavities in a commercial platform, Si wafers covered by a thick $SiO_2$ layer, acting as a membrane. We demonstrate that by exposing our material system to single ultrashort laser pulses above the evaporation threshold of Si, a controlled delamination of the cover layer without fracture can be achieved, leading to the formation of a microcavity accompanied by a surface bulging up to 1.5 µm. The formation dynamics of the microcavities has been investigated by time-resolved optical probe techniques. Ultrafast pump-probe microscopy resolves the initial stages of substrate melting, the onset of layer expansion followed by relaxation. Real-time reflectivity measurements with nanosecond resolution resolve the long-lived, highly complex oscillation process of the membrane, extending over tens of µs. The formed cavities can be re-excited at low fluence without damage or degradation, featuring vibrations at single frequencies with high Q factors. The frequencies can be tuned continuously from 8 MHz up to 19 MHz by adjusting the fabrication fluence. A simple classical model is found to well describe the behavior of the membranes. Our results pave the way for applications in micro- and nanooptics and acoustics, in great part due to the commercial availability of the Si-$SiO_2$ platform and the simplicity of the fabrication method. An extension to the field of microelectromechanical system (MEMS) of this technology is proposed.

## 1. Introduction

Silicon is arguably the most versatile and widely-spread material platform for countless technological applications in microelectronics and photonics[1], in great part due to its exceptional optical and electrical properties, combined with its high purity, crystal quality, abundance and low price. Although still rather small compared to the microelectronics market size (840 billion USD in 2024),[2] the silicon photonics market size has already reached 2 billion USD and is rapidly growing.[3] Within both sectors, short and ultrashort laser processing of silicon has long since become an indispensable tool for rapid wafer dicing,[4] surface texturing and volume inscription with nanometric features,[5] pulsed laser annealing after ion implantation,[6] as well as for the formation of amorphous surface layers with controlled thickness,[7] to name a few examples. There are important derivative platforms of standard silicon wafers, such as silicon on insulator (SOI), which consists of a silicon wafer with a buried $SiO_2$ layer. Its primary advantage is the reduction of parasitic capacitance and elimination of latch-up, making it ideal for high-speed, low-power designs of electronic circuits. Another derivative platform is silicon covered by a thermal oxide layer with a thickness of up to tens of micrometers (STO). Applications of STOs are mostly related to their excellent electrical isolation and passivation properties, which makes them ideal, for instance, for use in integrated circuits, optoelectronics and power devices.

Here, we introduce a novel application of STO wafers, the fabrication of vibrating microcavities confined by the silicon substrate and a bulged $SiO_2$ cover layer. The fabrication method is based on exposing a STO wafer to single ultrashort laser pulses above the evaporation threshold of Si, leading to a controlled delamination of the $SiO_2$ layer without fracture. The fabrication concept has already been reported by two groups. Serrano et al.[8] reported blister formation upon ns laser irradiation of STO wafers with oxide thickness below 100 nm using sub-10 µm spot sizes. McDonald et al.[9], reported blister formation upon fs laser irradiation of STO wafers with thicker oxide layers under similar conditions as ours, studying the surface topography as well as optical interference effects. Both works provided the proof-of-concept of fabricating spots with bulged cover layers without damage, but did not identify nor characterize their dynamic vibration properties during or after formation. Very recently, these properties have been reported in a different material system (thin metal films on transparent substrates).[10] In that work, the authors sent a focused femtosecond laser pulse through the back of the substrate, heating and partially melting the film, which expands. This causes partial delamination and thus the formation of so-called "femtosecond-laser delamination (FLD) cavities", featuring heights of several hundreds of nanometers. The authors beautifully demonstrated that vibrations of these cavities can be excited by ultrashort laser pulses, featuring high-Q acoustic resonances and coherent phonon modes in the above 100 GHz frequency range that last a few nanoseconds. The FLD cavities in STO we report here achieve heights of up to 1.5 µm, feature high-Q resonances in the MHz frequency range and extraordinarily long vibration life-times (tens of µs).

From a fundamental point of view, an interesting question is which are the underlying mechanisms that lead to the formation of the FLD microcavities in STO wafers. We have investigated in detail these mechanisms by employing two complementary time-resolved optical probe techniques. Ultrafast pump-probe microscopy[11],[12],[13] (UPPM) with fs temporal resolution upon fs laser irradiation of semiconductors is a powerful tool to identify the presence and duration of the multiple processes involved in similar situations, including free electron generation, lattice heating, thermal and non-thermal melting, and ablation. In turn, real-time reflectivity measurements (RTRM) with nanosecond resolution are able to monitor processes that last longer, including solidification via recrystallization or amorphization after short or ultrashort pulsed laser melting.[14] Its virtually unlimited temporal window is ideal for resolving the long-lived acoustic vibration of FLD cavities in STO.

It is worth pointing out that the here presented microcavities belong to the field of microoptomechanical systems (MOMS). which is a subclass of Microoptoelectromechanical systems (MOEMS),[15] a term coined by M.E. Motamedi.[16] Although our system is so far limited to optical excitation and response, an extension towards electrical excitation and response seems straight-forward by deposition of electrodes on both sides of the processed Si wafer, thus opening the possibility of a wider range of applications, including for instance capacitive pressure sensors.[17,18]

## 2. Results

This section is divided into five parts. In section 2.1, we will provide experimental evidence of laser-fabricated microcavities in silicon wafers covered by a micrometer-thick $SiO_2$ layer and show experimental results of the characterization of their spatial dimensions. We will also demonstrate tunability of their size and characterize the crystal structure of the Si wafer underneath. In section 2.2, the initial stages of cavity formation will be resolved by ultrafast optical pump-probe microscopy, reaching a maximum measurement window of 37 ns with a temporal resolution of 100 fs. In section 2.3 this time window will be expanded up to tens of microseconds by time-resolved point-probing reflectivity measurements with nanosecond resolution. Section 2.4 shows and discusses the results of re-excitation experiments of pre-fabricated microcavities in order to quantify their natural vibration frequencies. Finally, section 2.5 introduces a simple model to calculate the vibration frequencies of circular membranes and compares the results to the experimental data.

### 2.1. Fabrication and characterization of microcavities

The strategy for the fabrication of FLD microcavities in STO samples (Si wafer covered by 1.0 µm thick $SiO_2$ layer), as well as the underlying basic formation mechanism is illustrated in Fig. 1(a), the detailed experimental configuration being shown in Fig. S1 of the supplementary information. A single femtosecond laser pulse, ($\lambda$ = 800 nm, $\tau$ = 120 fs) is focused to a spot size of a few tens of micrometers onto the sample surface. Since the cover layer is transparent to the laser irradiation, all energy is deposited in a shallow surface layer of the Si wafer, causing melting and – for sufficiently high fluences – material evaporation. We propose that the high pressure induced by the evaporation process exerts strong forces onto the cover layer, causing it to deform, separate from the wafer and bulge, giving rise to a dome-like structure with a microcavity enclosed. We will support the claim of silicon evaporation to be the responsible mechanism for delamination later in this section by means of threshold experiments, as well as in sections 2.2 and 2.3 by means of time-resolved measurements.

Fig. 1(b) shows a topography map measured by AFM of such a microcavity formed by single pulse irradiation. As can be seen, the dome structure has radial symmetry. The off-center measurement was deliberate due to the limited lateral displacement range of the AFM motors (50 µm) in order to show the flatness of the sample beyond the extension of the dome together with the smoothness of the dome surface. This can also be appreciated in the horizontal height profile across the dome center, drawn along the green dashed line and plotted in Fig. 1(c). Concerning the shape of the profile, we have found that it can be well fitted by a cropped Gaussian function, although the underlying mechanisms responsible for deformation most likely follow other spatial dependencies, as discussed in detail in Ref. [19] . The dome features a maximum height of $h$ = 1.54 µm and a radius of $r$ = 25.2 µm. While the corresponding aspect ratio is rather small ($h/r$ = 0.06), the absolute height is very large compared to other types of structures with height elevations that have been reported in inorganic materials upon single pulse laser irradiation. For instance, Puerto et al. have reported surface swelling of phosphate glass through a laser-induced change of the complex refractive index, reaching a maximum amplitude of $h$ = 25 nm[20]. In some organic material, much larger heights can be achieved through a photochemical reaction process, as demonstrated by Bernabeu et al. in PVA/AA, reporting maximum height elevations of $h$ = 8 µm.[21]

In our case, the origin of height elevation is the deformation, separation and bulging of the cover layer, leaving underneath a microcavity. Fig. S2(a) of the supplementary information shows a SEM image of a cross section of a laser-fabricated microcavity cavity, confirming a partial delamination of the oxide layer. Importantly, delamination can also be identified and quantified non-destructively, using an optical microscope with monochromatic illumination. As sketched in Fig. 1(d), optical interference occurs between the light reflected at the two interfaces (air/cover layer and cover layer/cavity) of the bulged cover layer with the light reflected from the cavity/silicon interface. Due to the dome structure, the height differs for different spatial positions, which results in the appearance of circular constructive and destructive interference effects, so-called Newton rings [20],[22]. Fig. 1(e) displays an optical microscopy image recorded for illumination at $\lambda_{LED1}$ = 460 nm. The large circular central region, limited by a narrow dark ring whose diameter matches exactly the diameter of elevation measured by AFM, displays a considerable number of alternating bright and dark rings (Newton rings). The demonstration that interference is indeed the origin of this ring system is made by comparison to the microscopy image of the same dome recorded for illumination at $\lambda_{LED2}$ = 810 nm (c.f. Fig. 1(f)). For this longer illumination wavelength, the number of rings is approximately half of that at $\lambda_{LED1}$, as a consequence of the change of interference conditions. Assuming the cavity height to be equal to the dome height, the number # of rings (of both bright ones (constructive) and dark ones (destructive)) scales with the illumination wavelength as $\# = 4 \cdot h / \lambda_{probe}$,[23] which yields for $h$ = 1.54 µm a value of $\# \approx 13$ for $\lambda_{probe} = \lambda_{LED1}$ = 460 nm and $\# \approx 8$ $\lambda_{probe} = \lambda_{LED2}$ = 810 nm. These calculated values match perfectly the number of rings (including the central disk) that appear in Figs. 1(e,f), which proves the existence of a microcavity below the dome structure. The wavelength-dependent interference effect can also be appreciated in the image recorded of the same spot using white light illumination (see inset), which leads to colorful Newton rings.

We have modelled the reflectivity response of the system considering the experimentally measured topography and refractive index of the deformed cover layer as shown in Fig. 1(c). As shown in the Supporting Information (c.f. Fig. S3), the calculated reflectivity images match very well the experimental reflectivity images shown in Fig. 1(e,f). This fact corroborates that also quantitatively, monochromatic optical microscopy can be used to quantify the topography of the microcavity.

Noteworthy, the images shown in Figs. 1(e,f) also show an outer ring system, outside the narrow dark ring that marks the delimitation of the cavity. Specifically for Fig. 1(e), it is composed of three rings; an outmost narrow bright ring, followed by a narrow slightly less-bright ring and a wide bright inner ring. This ring system corresponds to a laser-amorphized region, in which a shallow amorphous surface layer has formed upon ultrafast melting and sufficiently rapid solidification to prevent the formation of the crystalline phase.[24] The optical response, consisting of alternating rings of increased reflectivity, has been widely investigated[7],[25] and is caused by the position-dependent thickness of the amorphous layer, giving rise to thickness-dependent interference effects (between the air/amorphous layer interface and the amorphous layer/bulk crystalline interface). The presence of interference effects in the amorphous region is confirmed by comparing the image to the one recorded at $\lambda_{LED2}$ = 810 nm, strongly changing the number and widths of the rings.

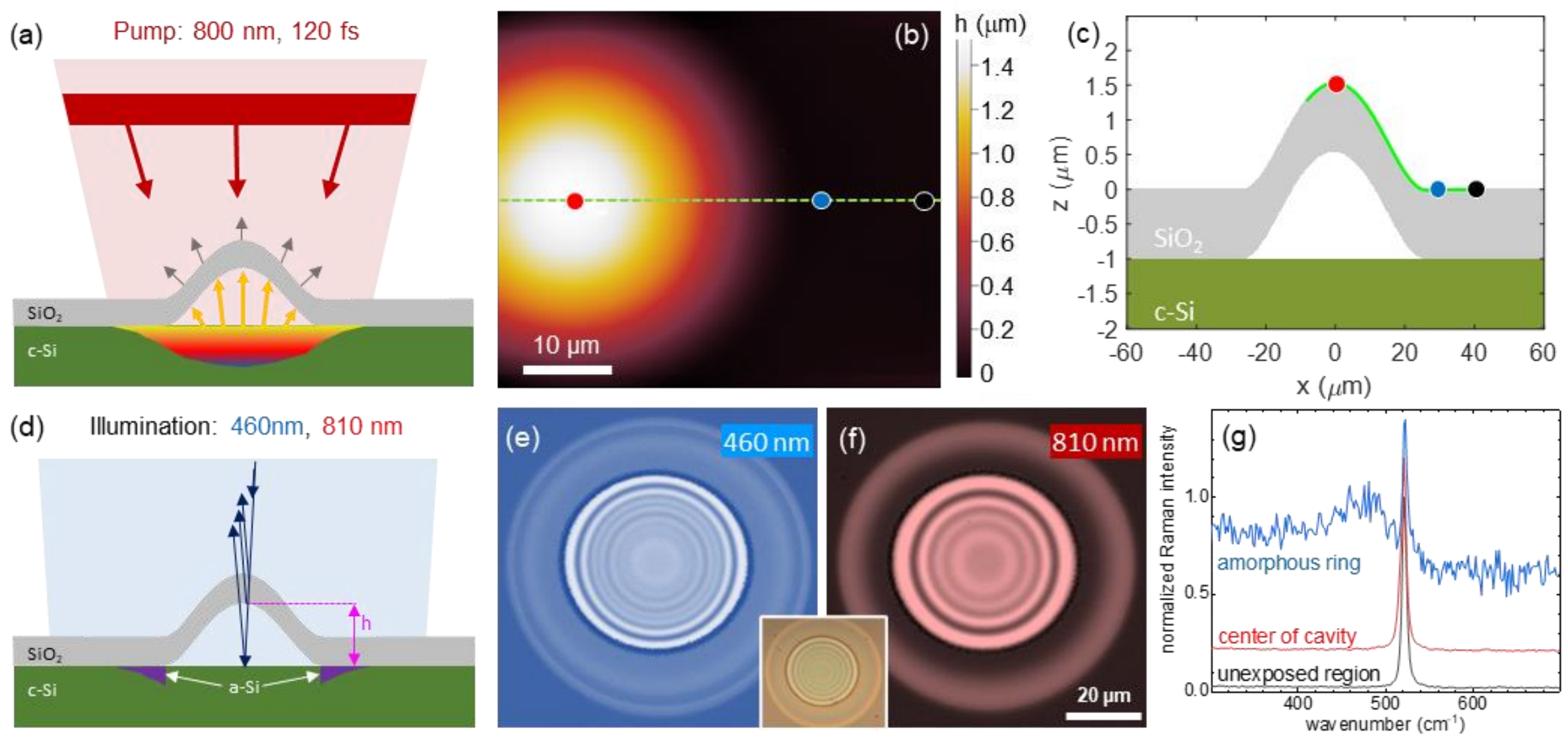

*Figure 1: (a) Sketch of the fabrication principle of optical microcavities based on focusing a fs pump pulse on a Si substrate with a $SiO_2$ cover layer, leading to heating, melting and evaporation of Si and detachment of the oxide layer. (b) Topography map of a microcavity fabrication by a single pulse at F = 603 ± 30 mJ/cm$^2$ measured by AFM. The green line and red, blue, black symbols mark specific spatial positions, used in (c) and (g). (c) True scale cross section of the microcavity determined from the AFM measurements. The green line corresponds to the height profile through the cavity center, extracted along the green line in the map displayed in (b). (d) Sketch of the final cavity cross section origin, giving rise to optical interference effects upon illumination with monochromatic light. (e,f) Optical microscopy images of the same cavity as measured by AFM in (b) using $\lambda_{LED1}$ = 460 nm and $\lambda_{LED2}$ = 810 nm as illumination wavelengths, respectively, revealing Newton rings caused by interference. The images have been recorded with a monochrome camera and are displayed in false colour scale to illustrate the illumination wavelength. The inset shows a true color image recorded with a color camera of the same spot using white light illumination, which leads to colorful Newton rings. (g) Micro-Raman spectra recorded at different positions, indicated in (b,c).*

The structural confirmation that this outer ring region is indeed amorphous is provided by micro-Raman spectroscopy measurements, displayed in Fig. 1(g). The spectrum recorded in this region displays the characteristic shoulder of the broad band centered at 473 cm$^{-1}$, which is the fingerprint of amorphous silicon. We have also recorded a spectrum in the unexposed region, featuring the characteristic strong peak of crystalline silicon, centered at 520 cm$^{-1}$. Interestingly, the spectrum recorded within the central region, containing the cavity, shows no evidence of the amorphous phase. This points towards a slowing-down of the solidification process as a consequence of the heat confinement of the evaporating silicon by the membrane, preventing ablation cooling[26] and allowing re-crystallization.

As a next step, we have investigated the influence of the laser peak fluence *F* on the width and height of the cavities. As detailed in the experimental section, the fluence is defined as $F = 2 \cdot E/(\pi \cdot \omega_x \cdot \omega_y)$, where E is the pulse energy and $\omega_x$, $\omega_y$ are the vertical and horizontal beam waist values, respectively. We assign an error of ± 5% to all fluence values quoted in this paper, mainly due to pulse-to-pulse fluctuations and long-term fluctuations of the laser system used, which will be lower for more modern laser systems. Fig. 2(a) displays a series of selected optical microscopy images of single laser pulse irradiations at different fluences, recorded upon illumination at $\lambda_{LED2}$ = 810 nm. In the low fluence region (170 mJ/cm$^2$ ≤ *F* ≤ 348 mJ/cm$^2$), only surface amorphization is observed, as can be appreciated by the progressive appearance of a characteristic three-ring structure of moderate reflectivity increase as a consequence of a progressive thickness increase of the amorphous surface layer. At a slightly higher fluence (375 mJ/cm$^2$), a small circular central region of high-reflectivity is formed in the center, which already corresponds to a cavity since it leads to constructive interference. As the fluence increases further, the cavity diameter increases and more and more interference rings appear as a consequence of the height increase.

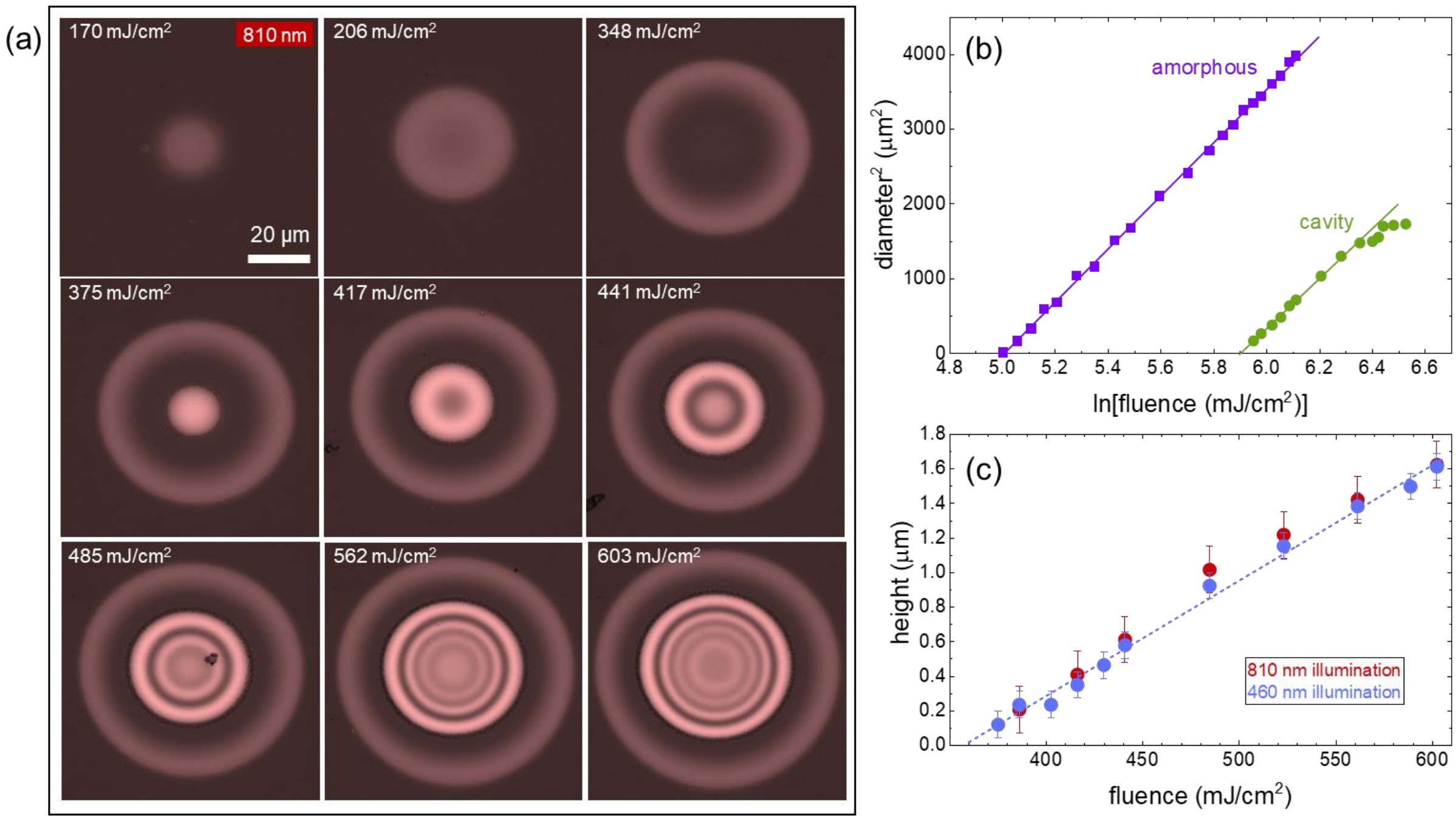

*Figure 2: (a) Optical microscopy images of a series of microcavities fabricated by single laser pulses at increasing fluences (specified by labels), using $\lambda_{LED2}$ = 810 nm as illumination wavelength. All fluence values have a determination error of ± 5%. (b) Fluence dependence of the square of the diameter of the outer amorphous region (purple symbols) and of the microcavity region (green symbols). The maximum experimental error in the diameter determined by optical microscopy is $\Delta d$ = ± 0.5 µm, which lies within the symbol size. (c) Fluence dependence of the cavity height h, calculated for $\lambda_{LED1}$ = 460 nm and $\lambda_{LED2}$ = 810 nm, by counting the number of interference rings # and using the expression $h = \# \cdot \lambda / 4$, as detailed in the text. The dashed curve is a linear fit to the 460 nm data.*

The plot in Fig. 2(b) shows a representation of the square of the diameter of the outer amorphous region $d_{am}^2$ (purple symbols) as a function of logarithm of the pulse fluence. This representation allows to evaluate to which extent the spatial laser intensity distribution at the sample plane is Gaussian, as proposed by Liu [27]. The fact that the data lie on a straight line over a wide fluence range confirms this assumption. This representation also allows to precisely determine the amorphization threshold by taking the intersection of the fitted line with the x-axis (= diameter equals zero), yielding $F_{am}$ = 150 ± 8 mJ/cm$^2$. This threshold value is in excellent agreement with other experimental studies on Si(111), covered by the native 3 nm-thick $SiO_2$ layer, which further underlines that the amorphization process is dictated by rapid melt-quenching [28].

Also included in Fig. 2(b) is a plot of the square of the diameter of the microcavity region $d_{cav}^2$ (green symbols). Also here, the data points can be fitted to a straight line, whose intersection with the x-axis yields the threshold fluence for cavity formation $F_{cav}$ = 362 ± 18 mJ/cm$^2$. This value is significantly above the measured ablation threshold for Si(111) covered by the native 3 nm thick $SiO_2$ layer ($F_{abl}$ = 289 ± 14 mJ/cm$^2$, measured in a separate experiment and consistent with the data reported in literature[28]). The higher threshold further supports the above-made claim of evaporation pressure being the origin of cavity formation. Additional evidence for evaporation being the main driving force for cavity formation has been obtained by performing TEM measurements of a cavity cross section, near the air/Si substrate. The results shown in Fig. S4 clearly identify the presence of a redeposition layer formed by Si nanoparticles on top of the Si substrate, the latter showing the presence of an ≈ 170 nm thick polycrystalline layer formed after solidification. Although the preparation of the cavity cross section by focused ion beam irradiation did not allow to include the fragile and detached 1 µm thick membrane in the cross section, we expect a similar Si nanoparticle layer to be present at the backside of the membrane.

We have also extracted the cavity height $h$ from each microscopy image of the cavities, calculated by counting the number of interference rings # (of both bright ones (constructive) and dark ones (destructive)), using the expression $h = \# \cdot \lambda_{probe} / 4$, as explained in Ref. [23]. Using this method, the error in the height determination can be specified as the uncertainty of identifying the maximum or minimum or being only close to it, which can be written as $\Delta h = \pm \lambda_{probe} / 6$. The result is plotted in Fig. 2(c) as a function of fluence, showing an approximately linear increase ($h = slope \cdot (F_0 - F_{cav})$ *for* $F > F_{cav}$) behavior up to a maximum height of $h$ = 1.55 µm for $F_{max}$ = 603 ± 30 mJ/cm$^2$. The observed behaviors under Gaussian beam irradiation demonstrates a deterministic fabrication process of the cavities with controlled tuneability of their dimensions by adjusting the peak intensity value with respect to the threshold. For higher fluences, partial or complete fracture of the cavity is observed (c.f. Fig. S5).

## 2.2. Formation process dynamics of elliptical microcavities: Initial stages

In this section, we have investigated the dynamics of the cavity formation during the first tens of nanoseconds. To this end, ultrafast optical pump-probe microscopy is the ideal technique [11],[12],[13] as it combines optical microscopy, the same technique used in the previous section for the determination of the 3D cavity structure, with a temporal resolution of better than 200 fs. While the detailed experimental configuration is shown in Section 4 (Experimental), the simplified sketch displayed in Fig. 3(a) shows the slightly modified excitation scheme with respect to section 2.1 (c.f. Fig. S1) together with the probing/detection configuration. Here, the circular 800 nm fs pump pulse is incident at an angle of $\theta = 52^o$, which correspondingly leads to an elliptical laser-excited region and final elliptical cavity. The 400 nm fs probe pulse is incident at $\theta = 0^o$ and illuminates the sample at different time delays t after the arrival of the pump pulse. The reflected portion of the probe light travels towards the optical microscope, which allows to form reflectivity snapshots of the sample surface at different time delays.

Fig. 3(b) shows selected snapshots corresponding to characteristic time scales for a selected fluence of $F$ = 422 ± 21 mJ/cm$^2$. At t = 0.5 ps, the strong and spatially homogeneous reflectivity increase (ΔR ≈ 40%) can be observed, which is the characteristic fingerprint of non-thermal melting in Si within a time much shorter than the time that electrons can transfer their energy to the lattice.[11] At t = 50 ps, the hot lattice has reached such a high temperature that the temperature-dependent reflectivity begins to decrease.[12] At t = 700 ps, an inner dark ring appears which evolves with time into the outer boundary of the cavity. At this stage, inner and outer regions are still highly reflective, which confirms the presence of the molten phase. At t = 2 ns, the outer bright ring contracts towards the spot center as a consequence of the onset of solidification, while the inner ellipse still remains homogeneous and featureless. However, at t = 5.3 ns, the inner bright region shows the first appearance of elliptical Newton rings, indicative of a detachment and elevation of the $SiO_2$ cover layer. As the time delay increases up to t =

23.2 ns, the number of the rings (and their optical contrast) increases strongly up to # = 12, which demonstrates a strong bulging of the oxide layer. Interestingly, for longer time delays, this tendency is inverted, strongly reducing the number of rings down # = 6 for the longest delay measured (t = 37.1 ns), which points towards a contraction of the cavity. Noteworthy, the ring number of the static final image recorded several seconds after the arrival of the pump is even lower (# = 5).

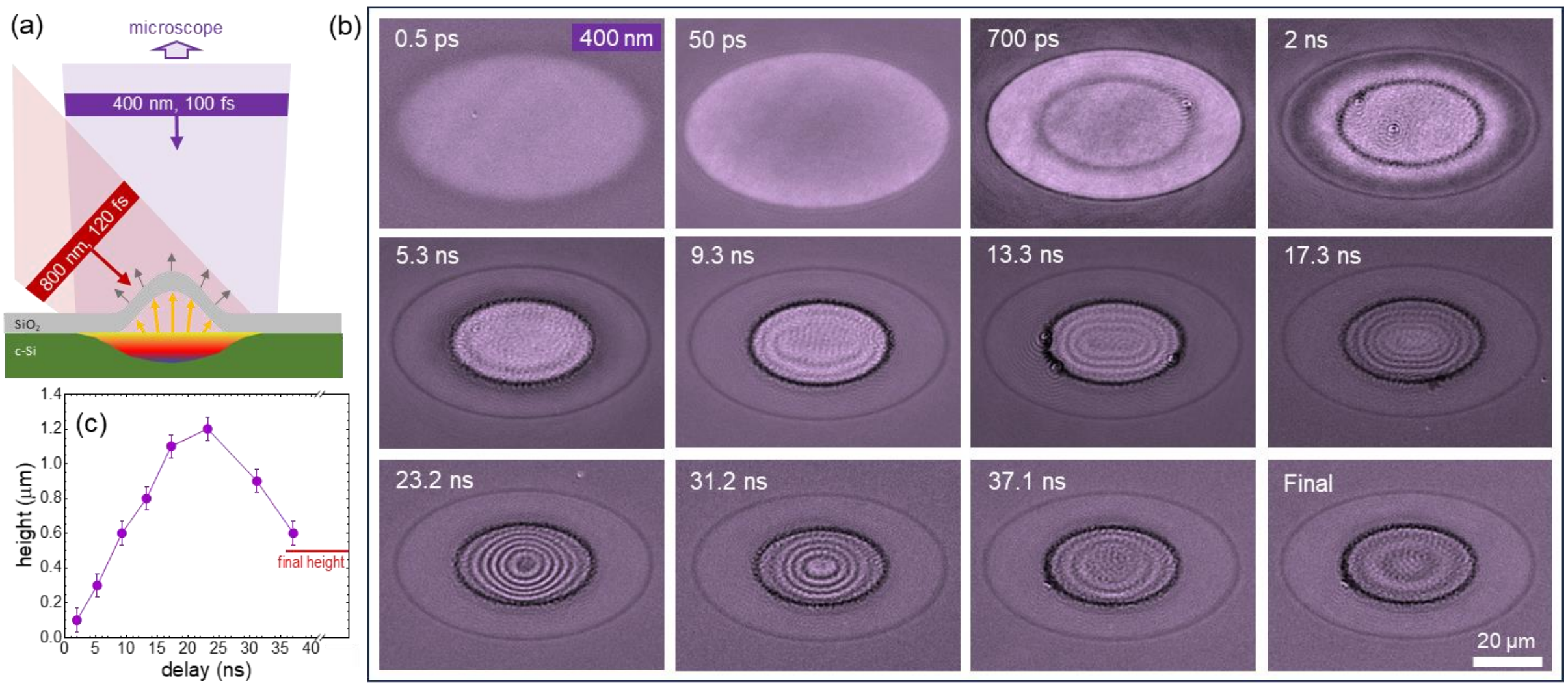

*Figure 3: (a) Sketch of the angled 800 nm fs laser excitation, delayed 400 nm normal probing scheme used for ultrafast pump-probe microscopy. (b) Reflectivity snapshots of the sample surface at different time delays t (indicated as labels) of the probe beam with respect to the pump beam ($F = 422 \pm 21$ mJ/cm$^2$). (c) Temporal evolution of the cavity height at the spot center determined from the ring number.*

Employing the same method as in section 2.1, we have extracted the cavity height $h$ for each snapshot, taking into account the different illumination wavelength used, $\lambda_{probe}$ = 400 nm. The temporal evolution of the height is shown in Fig. 3(c), featuring a threshold-like increase at t ≈ 3 ns with instantaneous expansion velocities as high as $v_{expand} = dh/dt$ = 75 m/s, before reaching a maximum height of $h_{max}$ ≈ 1.2 µm at t ≈ 23 ns. Afterwards, the cavity height decreases strongly, reaching a value of $h$ ≈ 600 nm at t ≈ 37 ns. The negative slope shown for the data in the regime of cavity contraction is considerable, corresponding to instantaneous velocities as high as $v_{contract} = dh/dt$ = 50 m/s.

The information on the expansion velocities can be used to investigate the physics of delamination/ablation thresholds, as it was extracted from a similar analysis of transient Newton rings during femtosecond laser ablation of atomically thin 2D-materials on glass substrate by Becher et al.,[29] reporting similar expansion velocities at fluences near the ablation threshold. The authors reported a film-thickness independent ablation threshold, which they attributed to a thermoelastic, acoustic delamination process. We have performed preliminary experiments on cavity formation attempts in Si wafers with oxide layers of different thicknesses, revealing that the ablation threshold (cavity fracture) does depend considerably on the film thickness, which we attribute to the fundamentally different energy deposition scenario and film thicknesses involved: In our case, the energy is deposited in the underlying substrate, causing melting and evaporation, which we propose to generate enormous pressures on the layer, whereas in Ref. [29] the energy is deposited in the film itself. Moreover, the film thicknesses there investigated are much thinner, in the range of few nm up to a maximum of 12 nm.

Another relevant study in this context has been performed by Rapp et al.,[30] who investigated the ablation behavior (without achieving cavity formation) of 100 nm and 500 nm thick $SiO_2$ films on silicon upon fs laser pulse irradiation. Using ultrafast pump-probe microscopy the authors quantified expansion velocity values of 1800 m/s and 400 m/s, respectively. These 5 - 24 times higher velocity values are likely to be responsible, at least in part, for the absence of cavity formation in that study.

An interesting observation that can be made from our data in Fig. 3(c) is that the height value obtained at the maximum time delay accessible with the experimental configuration ($h_{t=37ns} \approx 600$ nm) is only slightly higher than the final height of the cavity ($h_{final} \approx 500$ nm). A look at the evolution of the height-versus-time curve suggests that it is unlikely for the cavity to stabilize immediately at the final height, without an oscillation-like behavior passing through lower heights. We have investigated this hypothesis in the next section using a different technique that allows accessing much longer temporal windows.

### 2.3. Formation process dynamics of elliptical microcavities: Vibrations in the microsecond range

Real-time reflectivity measurements (RTRM) represent a powerful and versatile approach to complement the analyses. While the detailed experimental configuration will be presented later (see Section 4 (Experimental)), the simplified sketch displayed in Fig. 4(a) shows a pump configuration identical to the previous case presented in section 2.2, but a slightly modified probing scheme. Here, the probe beam is not pulsed but continuous and is focused at normal incidence at the very center of the cavity formed by the pump laser. The reflected portion of the probe light travels towards a fast photodiode with ns temporal resolution, which allows to measure for a single pump pulse the complete reflectivity evolution of the spot center over time. While the drawback of this method is the lack of spatial resolution compared to ultrafast pump-probe microscopy, its widely adjustable temporal window for observation up to the millisecond range is ideal for measuring long-lasting phenomena such as oscillating optical cavities.

Fig. 4(b) shows the recorded reflectivity evolution for the same excitation conditions as investigated in Section 2.2 with ultrafast microscopy. Consistent with the result shown in Fig. 3(b), an instantaneous and strong increase of the reflectivity due to melting is observed. In particular, the high reflectivity region appearing in Fig. 4(b) shows strong oscillations over time, which is equivalent to spatially evolving Newton rings for which the central region alternates between maxima and minima. Solidification, marked by a sudden strong reflectivity decrease is observed to take place at $t \approx 20$ ns, after which reflectivity oscillations continue to occur, lasting well beyond the chosen temporal window of 170 ns and without a defined period.

In order to verify the consistency of the two optical probing methods, we have compared the cavity height values $h$ extracted from both data sets, considering the different illumination wavelength used $\lambda_{probe}$ = 400 nm (Fig. 3(c)) and $\lambda_{probe}$ = 532 nm (Fig. 4(b)). The result is displayed in Fig. 4(c). Up to a delay of t = 32.6 ns, an excellent agreement of the results of both data sets is observed. For longer delays, the height data determination by RTRM alone becomes ambiguous since the technique does not provide the information if a new oscillation is caused by an increase or decrease of the total ring number (i.e. if the cavity expands or contracts). We have therefore plotted two possible height evolutions for the delay range not covered by ultrafast microscopy, according to the measured oscillation number. Using additional criteria, such as continuity height movement due to inertia of motion suggests a continuation of contraction (solid diamonds) but the inverse process (expansion, as indicated by open diamonds) cannot be excluded with certainty. A different pump-probe microscopy configuration with the capability of reaching long temporal delays would be required to definitively answer this question [23],[31].

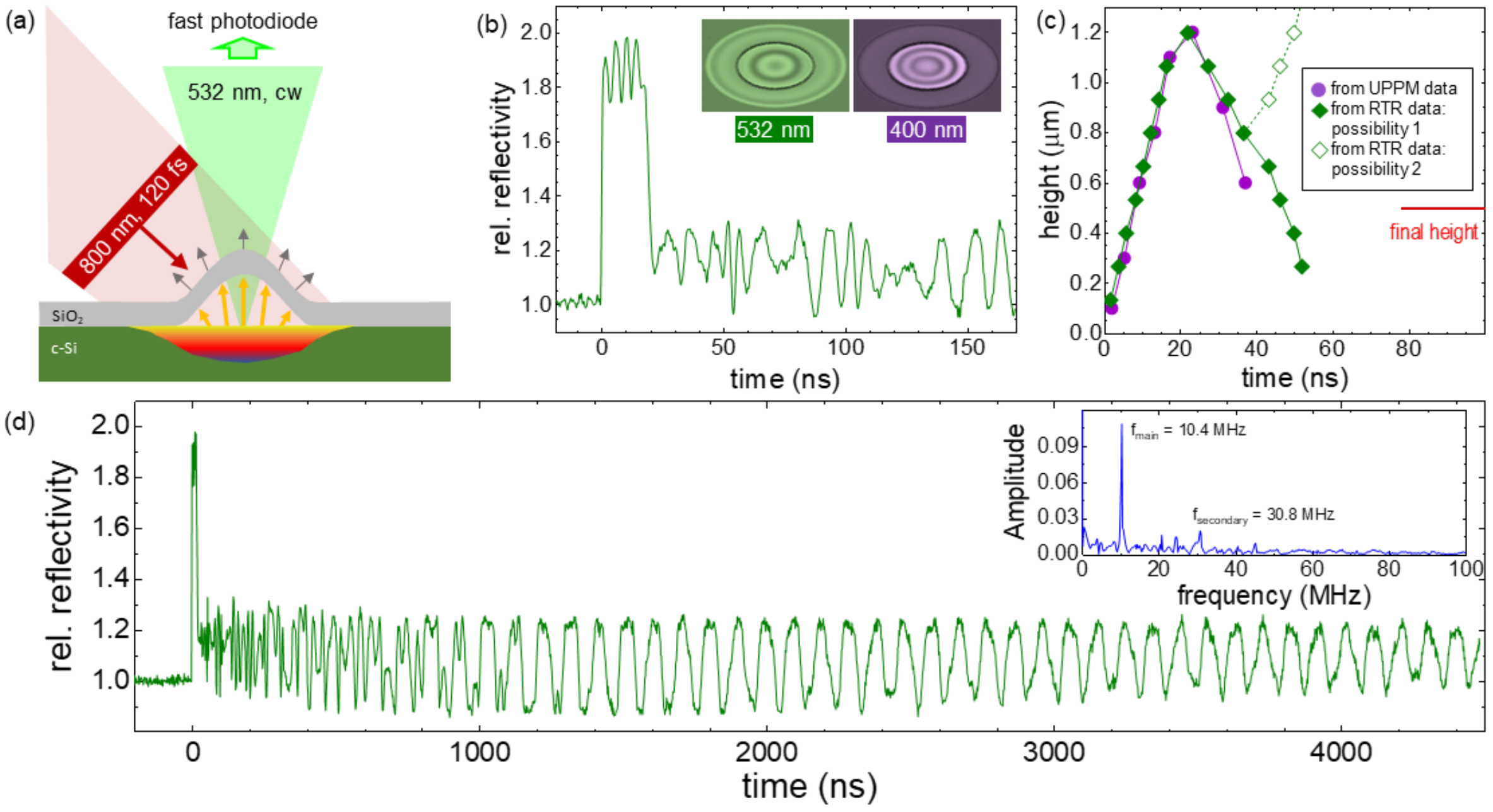

*Figure 4: (a) Sketch of the angled 800 nm fs laser excitation, continuous wave 532 nm normal and focused probing scheme used for real-time reflectivity measurements. (b) Reflectivity evolution during the first hundreds of nanoseconds of the sample surface upon irradiation at F = 422 ± 21 mJ/cm$^2$. The insets show the final appearance of the cavity upon illumination at $\lambda_{LED3}$ = 532 nm $\lambda_{LED4}$ = 400 nm in an optical microscope (c) Reconstructed temporal evolution of the cavity height determined from the reflectivity oscillation maxima and minima (green symbols) plotted together with the height data obtained from ultrafast microscopy (violet symbols, see also Fig. 3(c)). For better clarity, no error bars have been included, which are $\Delta h$(400 nm probe) = ± 67 nm and $\Delta h$(532 nm probe) = ± 89 nm. For delays t > 40 ns, two possible height evolutions are plotted due to the ambiguity of RTR measurements. (d) Reflectivity evolution over a longer time scale recorded under the same excitation and probing conditions as in (b). The inset shows the corresponding FFT of the reflectivity evolution, indicating the main frequency $f_{main}$ of the dominant oscillation, together with the second most intense frequency $f_{secondary}$. The maximum experimental error in the frequency determination is $\Delta f$ = ± 0.2 MHz.*

The fact that the membrane of the cavity vibrates during a very long time, much longer than the initial formation process can be seen in Fig. 4(d). Following the initial melting and solidification process, temporally resolved in Fig 4(b), aperiodic high-frequency oscillations take place up to approximately t = 1.25 µs. After that, the oscillations become periodic, are slowly attenuated and extend up to t > 15 µs (c.f. Fig. S6). It is worth noting that the transition from aperiodic to periodic oscillations corresponds to oscillations whose amplitude $A_{osci}$ does not exceed λ/4 of the probe wavelength, i.e. $A_{osci} \leq$ 133 nm, which corresponds to a reflectivity change $\Delta R_{osci} = R_{rel,max} - R_{rel,min} \approx 0.37$. The simple relation $A_{osci}(t) = \Delta R_{osci}(t) \cdot 133$ nm/0.37 allows scaling of the physical vibration amplitude at any moment for amplitudes below 133 nm. The inset in Fig. 4(d) displays a fast Fourier transformation spectrum of the reflectivity oscillations, featuring a main oscillation frequency of $f_{main}$ = 10.4 MHz, together with several weaker frequencies, amongst which the relatively strongest is $f_{secondary}$ = 30.8 MHz. In the next section, we investigate in more detail the relation between cavity properties (spatial dimensions and material properties) and the vibration frequencies.

### 2.4. Excitation of already formed cavities: Vibration frequencies of circular and elliptic micromembranes

The reflectivity oscillations shown in Fig. 4(d) show that they can essentially be grouped into a regime of cavity formation via laser delamination, followed by a complex damped oscillatory vibration process that ends in long-lasting highly periodic vibrations. Since the remarkable periodicity and long duration of the latter appear interesting for applications, it is desirable to find methods to re-excite vibrations in already formed cavities. To this end, we have explored two laser excitation strategies of existing cavities with laser fluences below the amorphization threshold, namely the one sketched in Fig. 4(a) and a new, collinear excitation strategy, sketched in Fig. 5(a). For both schemes, we have also tested re-excitation in a region nearby the cavity. In all cases, re-excitation of the characteristic main frequency $f_{main}$ was possible but only for direct excitation of the cavity region the signal-to-noise ratio was satisfying for fluences below the modification threshold. We have confirmed that several re-excitation experiments on the same cavity gave the same measured frequency. Moreover, switching the excitation laser system to nanoseconds allowed an unlimited number of re-excitation measurements without any sign of signal degradation.

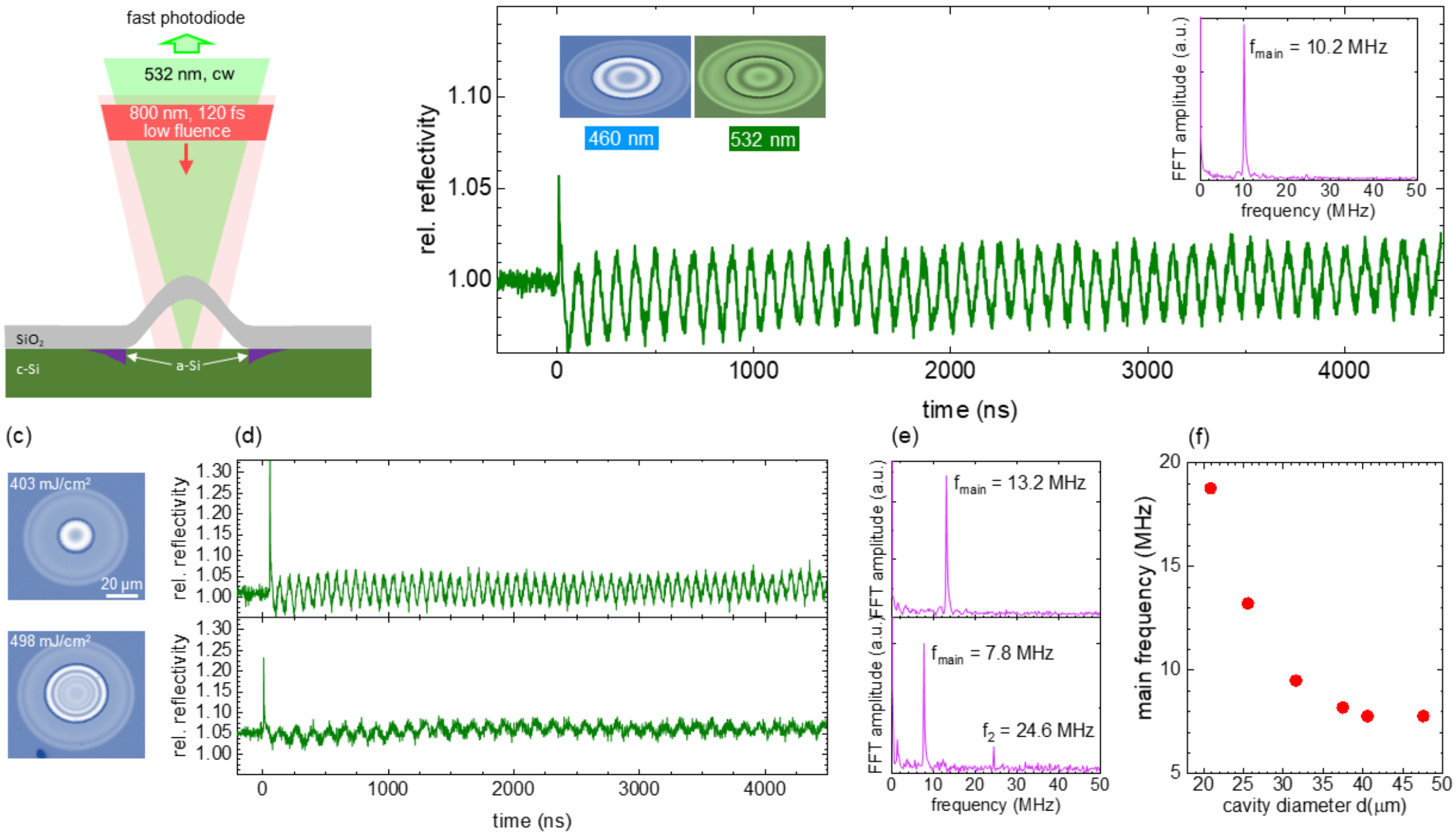

*Figure 5: (a) Sketch of the focused 800 nm fs laser re-excitation at normal incidence of prefabricated microcavities combined with collinear cw 532 nm focused probing used for real-time reflectivity measurements. (b) Reflectivity evolution of a prefabricated elliptical cavity as the one shown in Fig. 3 and Fig. 4, using colinear re-excitation at low fluences (below the modification threshold). Two insets show optical microscopy images at two different illumination wavelengths $\lambda_{LED1}$ = 460 nm $\lambda_{LED3}$ = 532 nm of the elliptical cavity. The third inset shows the corresponding FFT, indicating the main oscillation frequency $f_{main}$ (c) Optical microscopy image of two prefabricated circular cavities upon illumination $\lambda_{LED1}$ = 460 nm, with the indicated fluence values. (d) Corresponding reflectivity evolutions of the two*

*prefabricated cavities shown in (c), using colinear fs laser re-excitation at low fluences. (e) Corresponding FFTs calculated from (d), indicating the presence of two different main frequencies $f_{main}$ of the dominant oscillation and the presence of one secondary frequency. (f) Plot of $f_{main}$ as a function of the diameter of the laser-fabricated microcavities, determined from re-excitation measurements of six cavities whose diameters have been measured by optical microscopy. The maximum experimental error in the diameter determined by optical microscopy is $\Delta d = \pm 0.5$ µm and in the frequency $\Delta f = \pm 0.2$ MHz. Both lie within the symbol size of plot (f).*

Fig. 5(b) shows real time reflectivity measurement upon centered re-excitation at normal incidence of an identical elliptical cavity as the one studied in sections 2.2 and 2.3. The re-excitation fluence was chosen to lie well below the single pulse amorphization threshold to prevent damage. The reflectivity trace is dominated by highly periodic oscillations, lasting much longer than the selected time window. Yet, the particular time window was chosen to be best for determining the oscillation frequency by means of a FFT, displayed as inset in the figure. It can be clearly seen that the spectrum only contains a single frequency $f_{main}$ = 10.2 MHz, which is almost identical to the frequency observed for the late stages of the cavity formation process (c.f. Fig. 4(d)). This result demonstrates that fs-laser fabricated cavities have characteristic resonant frequencies that can be excited and detected optically at will. Noteworthy, the observed oscillation amplitudes are much smaller than those observed during cavity formation and have been calculated with the simple relation derived in the previous section as $A_{osci,re\text{-}ex} \approx 18$ nm.

The re-excitation strategy sketched in Fig. 5(a) also allows to study any type of microcavities, not only elliptical ones fabricated at 52º angle of incidence, but also circular ones. We have performed re-excitation studies on selected circular cavities fabricated at different fluences, equivalent to those shown in Fig. 2. Fig. 5(c) shows microscope images of two selected cavities for F = 403 mJ/cm$^2$ and F = 498 mJ/cm$^2$. The corresponding reflectivity traces are displayed in Fig. 5(d), showing long lasting periodic oscillations. It can immediately be seen that the oscillation periods are different from each other, as well as the oscillation amplitudes. Fig. 5(e) shows the corresponding FFT traces, yielding main frequencies $f_{main}$ = 13.2 MHz for the lower fluence, and $f_{main}$ = 7.8 MHz for the higher fluence. In the latter case, an additional high frequency of $f_2$ = 24.6 MHz appears, whose origin will be discussed next in the frame of a model that also explains the origin of the main frequencies.

It should be pointed out that the quality factor of these oscillations $Q = f_{main} / \Delta f_{main}$, with $\Delta f_{main}$ being the FWHM width, is rather high. For the two cases of circular cavities, the obtained values are $Q_{403mJ/cm2} \approx 47$ and $Q_{498mJ/cm2} \approx 33$. These values are limited by our current measurement technique using a fast photodiode and an oscilloscope, with a limited sampling rate and time resolution for a given measurement window, which leads to an underestimation of the obtained value. Yet, even despite this current limitation our results are almost as high as the quality factors obtained with higher resolution techniques on delaminated metal film membranes in Ref. [10], reporting values in the range of Q ≈ 40 – 60. We expect that a higher sampling technique applied to the cavities fabricated here would yield significantly higher Q factor values.

We have performed a total of six re-excitation measurements of cavities fabricated at different laser fluences. The obtained main oscillation frequencies are plotted in Fig. 5(f) as a function of the cavity diameter determined from optical microscopy. A strongly decreasing behavior with cavity diameter can be observed, starting with the highest frequency achievable under these conditions ($f_{main,max}$ = 18.8 MHz) and reaching a minimum value at $f_{main,min}$ = 7.8

MHz, with only small changes for large diameters. This result demonstrates that the oscillation frequency can be easily tuned over this range by a simple adjustment of the cavity diameter using different laser fluences.

## 2.5. Modelling vibration frequencies of circular membranes

We have developed a simple model to estimate the oscillation frequencies of microcavities from the dimensions of the laser-fabricated cavities and the properties of the $SiO_2$ layer membrane. We assume a flat circular membrane, clamped at a radius *r* that corresponds to half the cavity diameter *d* measured with optical microscopy, sketched in Fig. 6(a). In fact, the theory of such membranes with 2D vibration modes is derived directly from acoustics of drums as musical instruments[32,33,34]. Fig. 6(b) shows the first seven modes of vibration (*m,n*) of a circular membrane, where *m* denotes the number of nodal diameters and *n* the number of nodal circles. The ratio of modal frequency to frequency of the lowest mode is called the relative frequency $f_{rel}$. To convert these to actual frequencies, the following relation holds[35,36,37] :

$$f_{circ} = f_{rel} \times \frac{2.405}{2\pi r_{mem}} \sqrt{\frac{T}{\rho \cdot t}} \qquad (1)$$

where $r_{mem}$ is the membrane radius, $T$ is the tension in units of N/m, $\rho$ is the mass density $\rho = 2.2$ g/cm$^3$) and $t$ is the thickness of the membrane ($t = 1$ μm). Given the much larger diameter of the membrane than its height *h* (aspect ratios *d/h* range from 30 up to 110) in our case, we approximate $r_{mem} \approx r$. In our specific case of a clamped thin membrane, the tension $T$ is equivalent to the stiffness of the $SiO_2$ membrane,[38] which is a measure of the resistance to deformation in response to an applied force. Since this property depends not only on the material ($SiO_2$ in our case) but strongly on the size and shape of the membrane, we have opted to use it as a free parameter and discuss the obtained values.

The blue solid curve in Fig. 6(c) shows the calculated vibration frequency evolution of the membranes as a function of membrane radius, for the case of the fundamental oscillation mode (0,1) and using a constant value of T = 550 N/m for all diameters of the membranes fabricated. The agreement with the experimental data (red symbols) is astonishingly good for this very basic model, using a single value for the tension. This suggests that the classical concept of membrane vibrations holds even at this length and thickness scale. In order to explore the sensitivity of the result on the value of T, we have included in Fig. 6(c) the model result (blue dashed line) when using a lower value; e.g. T = 335 N/m, giving a lower bound of the tension value for obtaining a good match in the intermediate diameter range. The tension of the membrane is related to the stress $\sigma$ as $\sigma = T/t$. In order to verify that our obtained tension values are realistic, we have calculated the tension for typical residual stress values reported for thermally grown SiO2 films, which are $\sigma = 320$ MPa.[39] This yields T = 320 N/m, which is in excellent agreement with the obtained values from our fit. Noteworthy, the extracted tension value is considerably higher, but of the same order of magnitude, than experimentally measured tension values of sub-100 nm thick $SiO_2$ membranes of blisters in STO fabricated upon nanosecond laser irradiation[8].

While the model correctly describes a strong frequency increase for lower fluences (e.g. smaller membrane radii), the experimentally observed increase is even stronger. We have also considered the possibility that the measured frequencies do not correspond to the fundamental mode but to higher modes as those shown in Fig. 6(a). This possibility cannot be ruled out as shown in Fig. 6(c), displaying the curves for the (1,1) mode with two different values of *T*. Using *T* = 550 N/m for mode (1,1) predicts too high frequency values. However, using *T* = 215 N/m for mode (1,1), an identical curve as for T = 550 N/m for mode (0,1) is obtained, revealing an inherent ambiguity of vibration mode with surface tension. More complex modelling[37] and complementary experimental mode imaging is required to investigate in detail the frequency response and vibration modes present in this system.

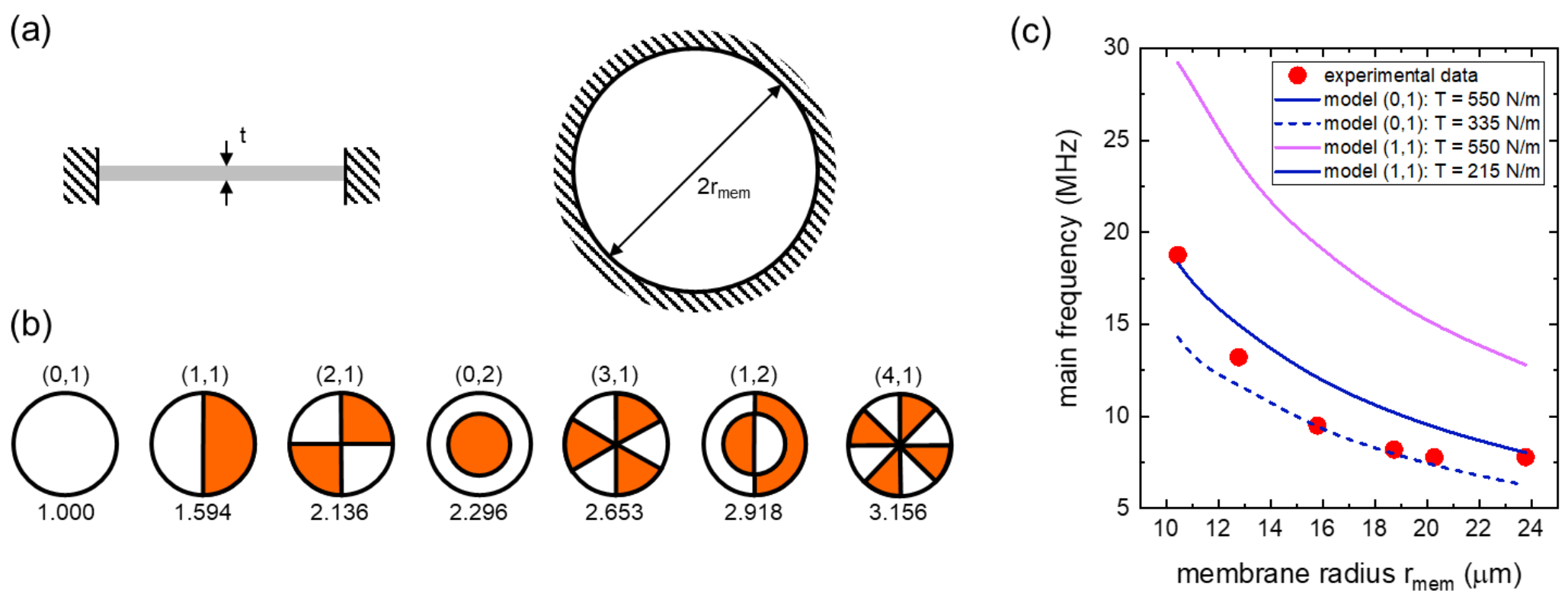

*Figure 6: (a) Sketch of side- and top-views a clamped membrane with thickness t and radius $r_{mem}$, which corresponds to the model system for the here-studied detached $SiO_2$ layer of the microcavity (b) First seven modes of vibration (m,n) of a circular membrane. m denotes the number of nodal diameters and n denotes the number of nodal circles. The corresponding relative frequency $f_{rel}$ is given below each mode. (c) Comparison of the obtained model results (curves) calculated for two vibration modes ((0,1) and (1,1)) and different tension values T to the experimentally measured oscillation frequencies (red symbols), taken from Fig. 5(f). The maximum experimental error in the cavity radius determined by optical microscopy is $\Delta r_{mem}$ = ± 0.25 µm and in the frequency $\Delta f$ = ± 0.2 MHz, both within the symbol size of plot (c).*

## 3. Conclusions and Outlook

We have developed a simple method to fabricate circular and elliptical microcavities on commercial Si-based platform (STO), Si wafers covered by a micrometer thick $SiO_2$ cover layer. It is based on controlled delamination of the cover layer without fracture upon single fs laser pulse exposure. The underlying mechanism relies on the high pressure generated by the evaporating silicon at high fluences, causing a bulging of the surface. The lateral dimensions and the height can be easily quantified by using optical microscopy with monochromatic light. The largest cavities fabricated under the present condition feature a maximum height of $h$ = 1.54 µm and a radius of $r$ = 25.2 µm and their surface curvature can be well fitted to a cropped Gaussian function.

Two complementary time-resolved optical probe techniques have been used to temporally resolve the formation dynamics of elliptical microcavities. Ultrafast pump-probe microscopy reaches out to a maximum delay of t = 37 ns and confirms the presence of the molten phase and quantifies the onset of layer expansion at t ≈ 2 ns, reaching a transient maximum height at t ≈ 23 ns after which it contracts again. Real-time reflectivity measurements with nanosecond resolution are able to monitor much longer time scales and reveal that complex, aperiodic height oscillations take place over a window of 1 µs, followed by periodic oscillations that last more than 15 µs.

We have also demonstrated that the membranes of the formed cavities, circular or elliptical and of any size and height, can be re-excited at low fluence without damage or degradation. They show long-lived oscillations with a single dominant frequency in the MHz range, characteristic to its size and shape. We have introduced a simple classical model of vibrating membranes with different vibration modes based on acousto-mechanics, which is able to reproduce the experimentally measured frequencies. We expect that the here presented fabrication strategy based on single pulse irradiation of an inexpensive and commercial material platform STO, together with the simple optical technique to measure the vibrations may find application in the fields of ultrasound (for instance as ultrasonic transducers), optical encoding and readout, mechanics, amongst others. We anticipate that there is a considerable number of other material systems, composed of thick optically transparent layers on top of absorbing substrates, that should have strong potential for cavity formation with vibration response upon excitation. Moreover, an extension to the field of microelectromechanical systems (MEMS) of this technology appears straightforward with a wide range of additional applications such as pressure sensors.

## 4. Experimental

The sample studied is a single-side polished, n-doped (111) Si wafer with a resistivity of 1 – 10 Ohm·cm and a thickness of 525 ± 25 µm, covered by 1.00 ± 0.05 µm thick $SiO_2$ layer grown by a wet thermal oxidation process, purchased from MicroChemicals Germany. Single laser pulse irradiations have been performed using a regeneratively amplified Titanium:Sapphire laser system (Spectra Physics, Spitfire Pro) operating at 800 nm, 120 fs pulse duration, with a maximum pulse energy of 1 mJ and operated at a repetition rate of 100 Hz. A simplified sketch of the irradiation configuration at normal incidence with respect to the sample surface is shown in Fig. S1 of the Supporting Information. Briefly, the pulse was sent through a shutter to select single pulses from a 100 Hz pulse train and the pulse energy was controlled by a combination of a motorized half-waveplate and two cascaded thin film polarizers (symbolized in the figure by a polarizing beam splitter cube). The emerging beam was demagnified with a 3x telescope and cleaned with a circular aperture of 2 mm in diameter, then focused at normal incidence ($\theta = 0^o$) with a f = 150 mm lens onto the sample surface. The intensity distribution at the sample surface was measured experimentally using single pulse imprints of amorphous marks as proposed by Liu [27], obtaining excellent fits for a circular Gaussian distribution, yielding a beam waist (1/e$^2$ radii) of $\omega_{x,y}$ = 47.6 µm. All fluence values quoted refer to the peak fluence, defined as $F = 2 \cdot E/(\pi \cdot \omega_x \cdot \omega_y)$, where E is the pulse energy and $\omega_x$, $\omega_y$ are the vertical and horizontal beam waist values, respectively. We assign an error of ± 5% to all fluence values quoted in this paper, dominated by the relatively high pulse-to-pulse and long-term fluctuations of the relatively old laser system used.

An in-situ optical microscope, composed of a 10x objective lens, a tube lens and a CCD camera (protected from laser light scatter by a short pass filter) was used for inspection of the irradiation area at an angle, illuminated by a white light source from the opposite side.

In order to investigate the transient changes of the sample upon fs laser excitation with temporal resolution, a more complex configuration needed to be developed. It consisted of a modified irradiation configuration and two different optical probe configurations implemented in a single setup, as shown in Fig. 7. The 800 nm laser beam was divided into a pump and probe beam by means of a beam splitter (not shown). The pump beam (lower branch in the figure) was sent through the shutter, followed by a circular aperture of 3.5 mm in diameter and a halfwave plate to select p-polarization, followed by an optical delay line (delay line 1). It was then focused with a f = 150 mm lens onto the sample surface at $\theta = 52^o$ angle with respect to the sample surface, leading to an elliptical intensity distribution at the sample surface. The horizontal and vertical beam waists (1/e$^2$ radii) have been determined experimentally as explained in the previous paragraph, yielding $\omega_x$ = 37.5 µm and $\omega_y$ = 21.8 µ, respectively.

In addition to the pump configuration, two optical probing schemes were added, addressing different time windows: First, an ultrafast pump-probe microscopy (UPPM) setup, sketched in Figure 7 and employed in similar configurations in several other studies Ref. [4012]. The probe pulses were obtained by frequency-doubling the upper branch of the 800 nm beam using a barium borate crystal (BBO) to obtain 400 nm, which was filtered from residual 800 nm light with a short pass filter. The probe beam was sent through a prism compressor in order to pre-compensate the group velocity dispersion (GVD) experienced by the pulse during its pass through the optical elements of the entire beam path up to the sample, ensuring a pulse duration of approximately 150 fs at the sample plane. At the exit of the prism compressor, the 400 nm probe beam was sent to a second delay line (delay line 2), composed of a translating a platform with two mirrors at 45°, which retroreflected the probe beam parallel to the incident beam towards a fixed end mirror. Translating the platform along a long rail, optical delay times as long as 37 ns could be reached with intermediate delay values as desired. The retroreflected probe beam was coupled into a home-built microscope by means of a quarter-wave plate and a polarizing beam splitter cube (PBS2), followed by a 2x magnification telescope and a lens to provide homogeneous sample illumination. The microscope was composed of a long-distance objective lens (20x, NA 0.42), a tube lens and a 12-bit CCD camera.

This setup allowed us to record images of the surface at different delays after the arrival of the pump pulse, offering sub-picosecond temporal and micrometer spatial resolution. The temporal delay between the pump and probe pulse was controlled using two optical delay lines. Delay line 1 (pump beam path) was motorized and covered a total range from 0 to 1 ns, in steps of 20 fs. Delay line 2 (probe beam path) was manual and could be set to any delay value in the range from 1 ns to 37 ns.

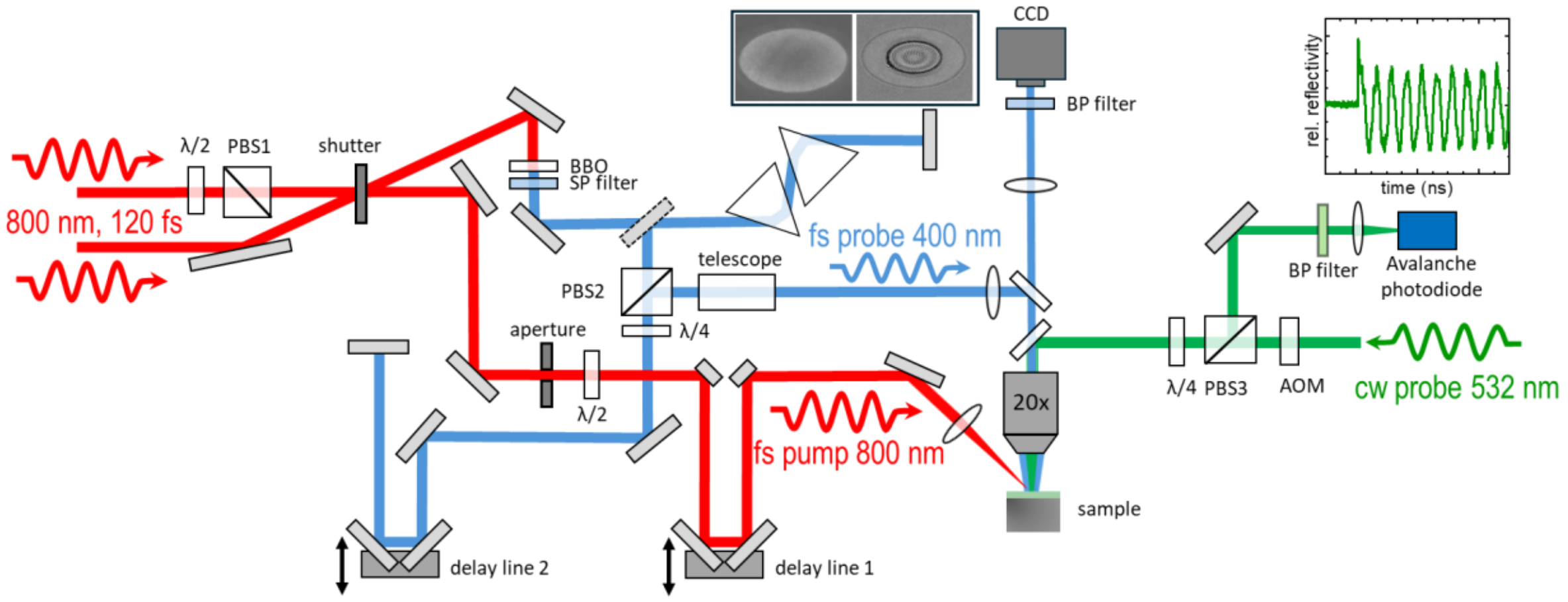

*Figure 7: Sketch of the 800 nm, 120 fs laser excitation setup (red beam), combined with a fs probe beam (blue beam) ultrafast reflection microscopy (UPPM) setup and a continuous wave 532 nm (green beam) real-time reflectivity measurement (RTRM) point probing setup. The acronyms shown in the figure correspond to half-wave plate (λ/2), quarter-wave plate (λ/4), polarizing beam splitter (PBS), short pass filter (SP filter), band pass filter (BP filter), charge coupled device camera (CCD), and acousto-optic modulator (AOM). See text for a detailed description of the setup.*

For acquisition of a series of time-delayed snapshots a non-irradiated surface area is used for each delay value and three different images are recorded at that delay value: Image 1, acquired before irradiation blocking the pump beam, is used for normalization purposes. Image 2, acquired at the desired pump-probe delay upon irradiation with the pump beam unblocked, is used to obtain the state of the surface during the ablation/modification process. Image 3, acquired at a very long delay (typically 1 second) after irradiation with the pump beam blocked, is used to obtain the state of the surface after irradiation. That way, for each delay value two normalized reflectivity 32-bit images were obtained; one corresponding to the chosen delay value (obtained by dividing Image 2 by Image 1) and one corresponding to the final state after irradiation (obtained by dividing Image 3 by Image 1). In these images, pixel values of 1.00 correspond to the reflectivity of the pristine material surface. Zero delay between pump and probe pulse has been defined by assigning to the first image with a weak reflectivity increase above noise level a delay of + 100 fs.

The second optical probe setup allowed for real-time reflectivity measurements upon fs laser excitation and is included in the right part of Figure 7. Briefly, the transient reflectivity changes occurring upon fs laser pulse irradiation were monitored using a cw probe beam and a fast photodiode, similar to the configuration reported in Refs. [4142]. The probe beam was a single-mode intra-cavity frequency-doubled Nd:YAG laser at 532 nm with horizontal polarization; the continuous wave beam was temporally modulated by means of an acousto-optical modulator to produce single rectangular pulses of 50 µs duration at the repetition rate of the fs laser (100 Hz), temporally synchronized. The probe pulse was injected into the in-situ microscope also used for pump-probe microscopy, being reflected by a beam splitter towards the sample and focused by the objective lens onto the sample at normal incidence to a circular spot with a measured beam waist of 6 µm at the center of the fs-laser excited area. This size is sufficiently small to probe even the smallest cavity investigated (c.f. Fig. 2(a)) without probing the surrounding region and sufficiently large to avoid to be affected by possibly present nodal diameters or circles of higher order oscillation modes, where the membrane displacement would be zero and no oscillations be detected.

The back-reflected probe beam was recollimated by the objective lens, rotated to vertical polarization by means of a quarter-wave plate ($\lambda/4$) in double pass, which causes it to be reflected by 90º inside PBS3, and was eventually focused onto a fast avalanche photodiode (APD). Unwanted collection of scattered pump light was blocked with a short pass filter. Figure 7 includes a real measurement of the transient reflectivity change occurring in an already formed microcavity upon single fs laser pulse re-excitation, as measured by the APD. Represented is the normalized relative reflectivity $R_{rel}(t)$, defined as $R_{rel}(t) = R(t)/R0$, where $R0$ is the initial sample reflectivity before pump pulse exposure and $R(t)$ the transient material reflectivity evolution upon irradiation, both measured with the APD. The maximum temporal resolution of the setup has been determined by employing the 10% - 90% rise time method upon measuring an ultrashort laser pulse, yielding a value of 600 ps for a temporal window of 50 ns. For larger temporal windows, the temporal resolution of the system is reduced in correspondence with the maximum number of data points of the oscilloscope, yielding 2.4 ns for a temporal window of 200 ns.

After irradiation, the sample was inspected by means of an optical microscope (Nikon Eclipse Ti), supplied with two monochromatic illumination LED light sources ($\lambda_{LED1}$ = 460 nm, $\lambda_{LED2}$ = 810 nm) and a tungsten-halogen white light source with narrowband filters allowing illumination at two additional wavelengths (($\lambda_{LED3}$ = 532 nm, $\lambda_{LED4}$ = 400 nm). Images were recorded with a 50x objective (NA = 0.8) and a 12-bit CCD. Due to the presence of the microcavity at the sample surface, the recorded images show optical interference effects in form of Newton rings. Optical modelling of the microcavities is carried out using multilayer modelling based on the transfer-matrix method (TMM), considering the refractive indices (n and k) of the substrate (c-Si), possible amorphization (a-Si), the top oxide cover layer ($SiO_2$), and air/vacuum when the oxide layer is delaminated. In this work, as previously introduced in Ref. [43], we design a 2D structure that accounts for both the in-depth layer distribution (z-axis) and the lateral distribution (x-axis). Therefore, this modelling approach enables us not only to simulate the experimental reflectivity profiles but also, by applying radial symmetry, to generate simulated images of the reflectivity behavior of the full laser-induced modification. This includes amorphization at the edges and a microcavity at the center, as shown in Figure S3 in the Supporting Information.

On selected microcavities, micro-Raman measurements (Renishaw Ramascope 2000) were performed in backscattering geometry with an Argon ion laser at 514.5 nm and a metallographic optical microscope (Olympus BH-2, 100 x Objective). Laser power was kept low to avoid any surface damage. The topography of selected microcavities has been characterized by atomic force microscopy (AFM) (Park Systems, XE7) in non-contact mode.

**Supporting Information**

- Sketch of the laser irradiation setup for fabrication of circular microcavities (Fig. S1)
- SEM image of a cross-section of a laser-fabricated microcavity (Fig. S2)
- 2D simulation of refractive index cross section of a microcavity, matching the shape of the AFM profile together with simulated reflectivity images at different wavelengths (Fig. S3)
- TEM image of a microcavity cross section, near the air/Si substrate (Fig. S4)
- Optical microscopy images of microcavities above the damage threshold (Fig. S5)
- Long-time reflectivity evolution recorded during the formation process of a microcavity (Fig. S6)

**Acknowledgements**

This work has been supported by grants PID2023-148178OB-C22, FPU23/01300, and Severo Ochoa Centers of Excellence program CEX2024-001445-S, funded by MICIU/AEI/10.13039/501100011033, as well as grant TEC-2024/TEC-85, funded by Consejería de Educacion, Ciencia y Universidades (Comunidad de Madrid, (Spain)).

**Author contributions**

G.G.-M.., M.G.-L., D.G. and J.Si. performed preliminary irradiation experiments that led to the discovery of the microcavities. J.Si. conceived the idea of studying the dynamics of cavity formation and re-excitation. B.V. and J.Si performed the corresponding experiments, including non-time-resolved irradiations, analyzed the data and implemented the theoretical model. I.S. performed the AFM measurements. F.A.-R. performed the Raman spectroscopy measurements. J.Si. and A.F.-R. shared the supervision of B.V. J.Si. and J.So. achieved the necessary funding and provided the experimental infrastructure for the research. All authors discussed the results, commented on the paper, and contributed to the writing of manuscript.

**Data availability**

The data that support the findings of this study are available within the paper and the Supporting Information. Additional data related to this paper are available from the corresponding authors upon reasonable request.

**Conflict of interests**

The authors declare no conflicts of interest.

# Supporting Information

Laser fabrication of vibrating microcavities on silicon:
Formation dynamics and frequency tunability

Belen Valiente[1,*] Gonzalo Gomez-Munoz[1], Mario Garcia-Lechuga[1], Irene Solana[1], Fernando Agulló-Rueda[2], David Grojo[3], Angel Rodriguez-Fernandez[4], Javier Solis[1], and Jan Siegel[1,*]

[1] Laser Processing Group, Instituto de Optica (IO), CSIC, Serrano 121, 28006 Madrid, Spain
[2] Instituto de Ciencia de Materiales de Madrid (ICMM), CSIC, 28049, Madrid, Spain
[3] Aix-Marseille Université, CNRS, LP3 UMR 7341, Marseille 13009, France
[4] Diamond Light Source, Harwell Science & Innovation Campus, Didcot, OX11 0DE, UK

[*]Corresponding author emails: belen.valiente@io.cfmac.csic.es, j.siegel@csic.es

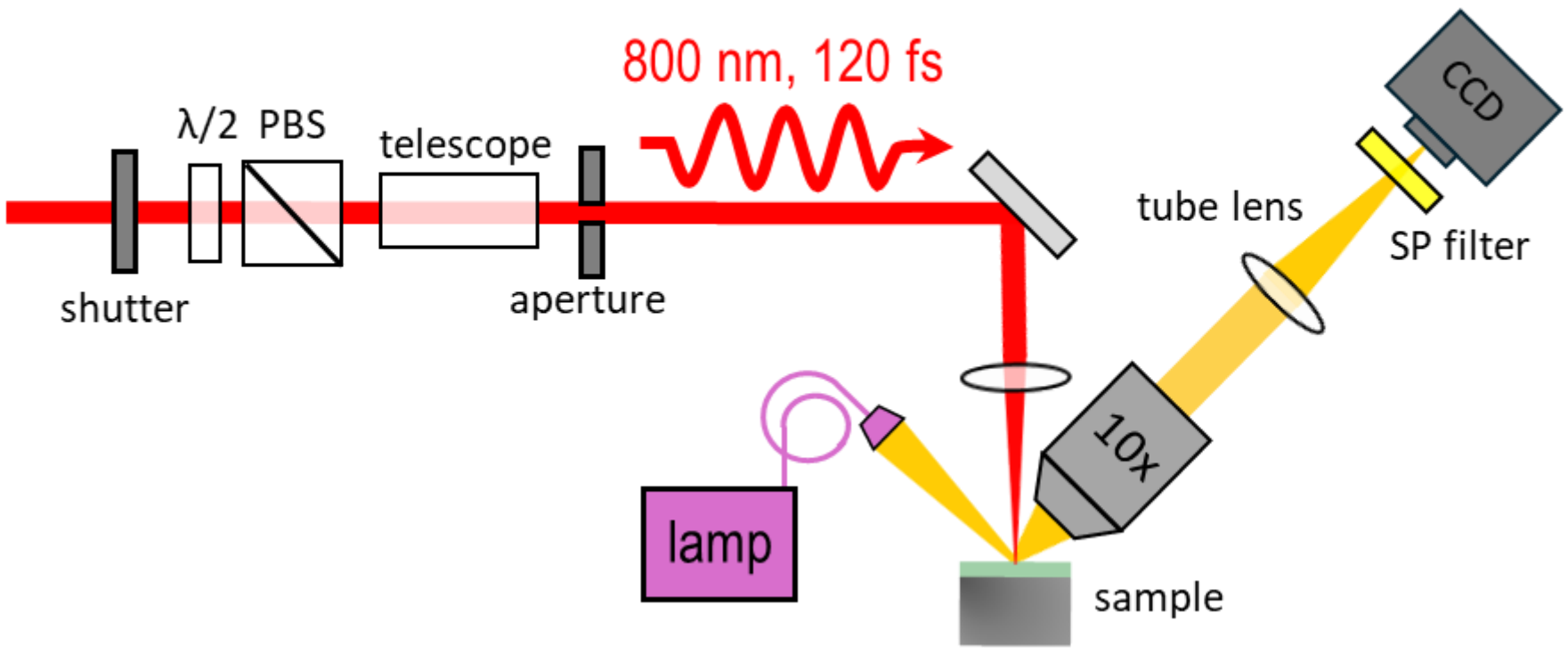

*Figure S1: Sketch of the 800 nm, 120 fs laser excitation setup (red beam), combined with an in-situ white light illumination and angled microscopy The acronyms used correspond to half-wave plate (λ/2), polarizing beam splitter (PBS), short pass filter (SP filter), charge coupled device camera (CCD). See text for a detailed description of the setup.*

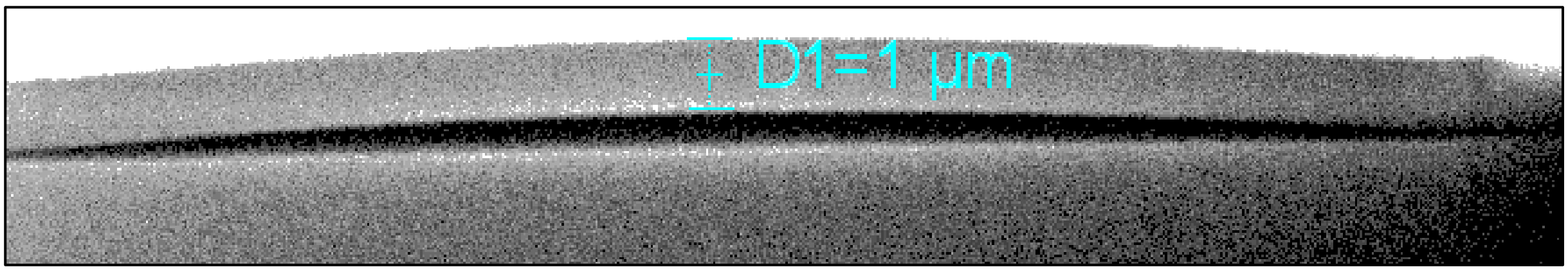

*Figure S2: SEM image of a cross-section through the center of a laser-fabricated microcavity prepared by focused ion beam milling. The vertical scale measurement corresponds to the oxide layer, separated by a gap (dark region) from the underlying silicon substrate.*

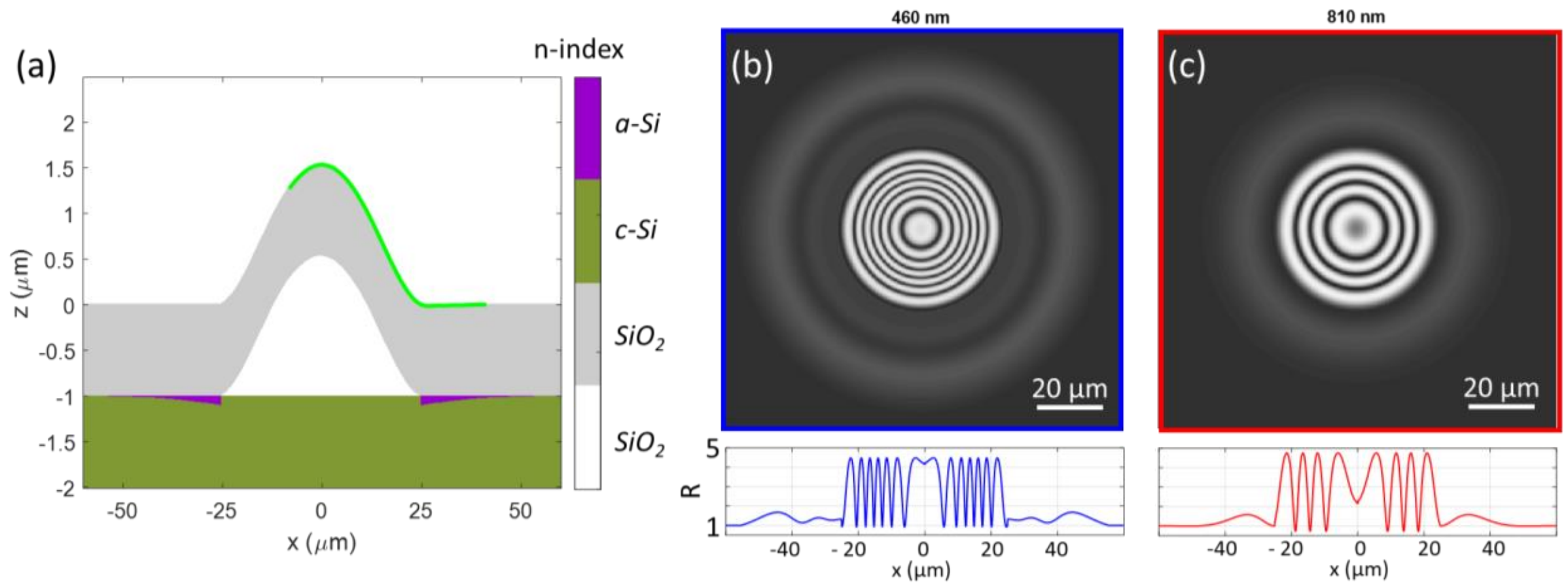

*Figure S3: (a) 2D refractive index mapping of the simulated laser-induced transformation obtained under F = 603 mJ/cm$^2$. The geometry of the $SiO_2$ elevation is computed to match the exact shape of the AFM profile (plotted in light green). (b–c) Simulated reflectivity images under illumination at 460 nm and 810 nm, respectively, corresponding to the 2D profile shown in (a). The images are generated by applying radial symmetry to the reflectivity profile shown underneath.*

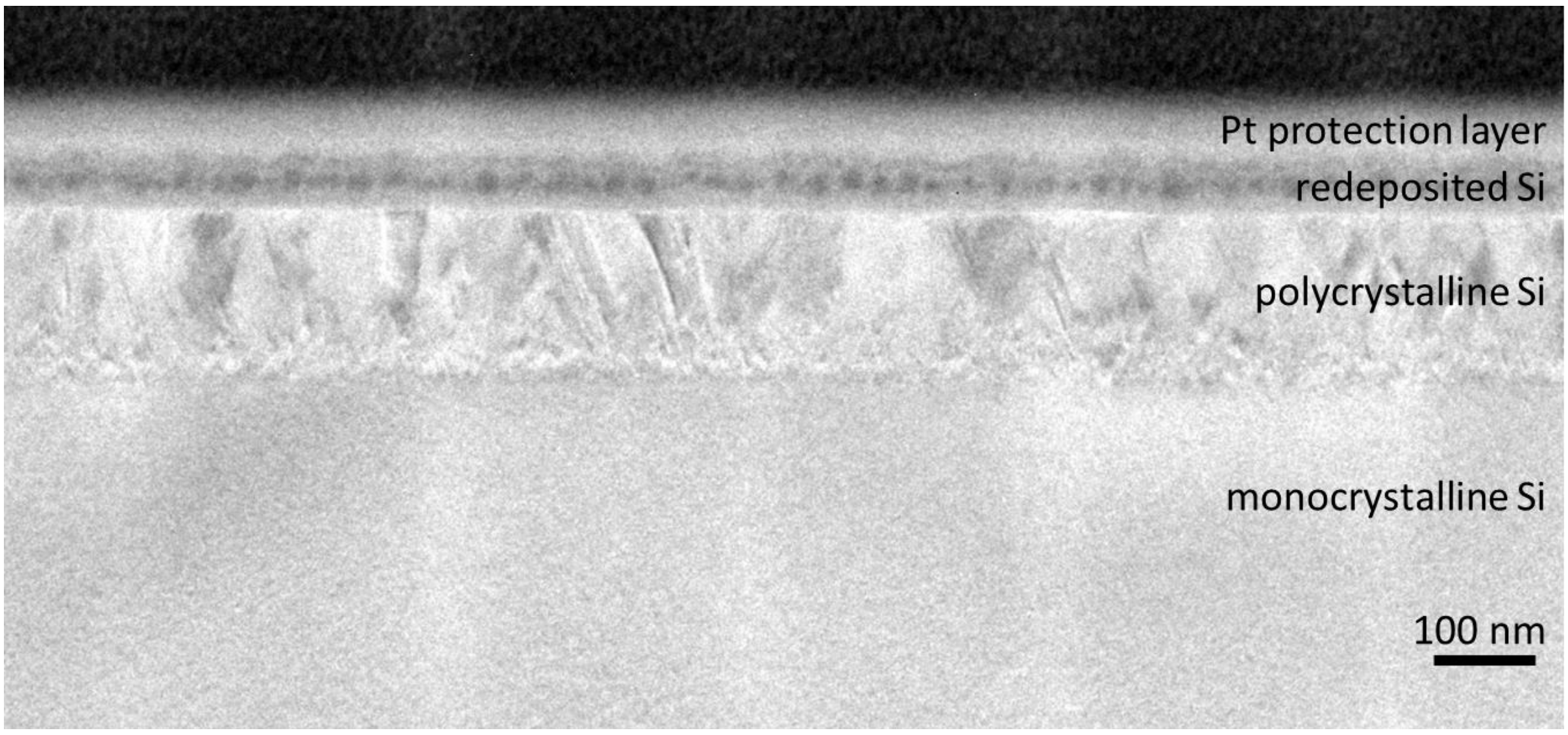

*Figure S4: TEM image of a cavity cross section, near the air/Si substrate (not showing the $SiO_2$ membrane), showing the different layers after melting, evaporation and solidification, revealing the presence of a redeposition layer formed by Si nanoparticles on top of a polycrystalline Si layer. The Pt layer is only due to the TEM preparation method.*

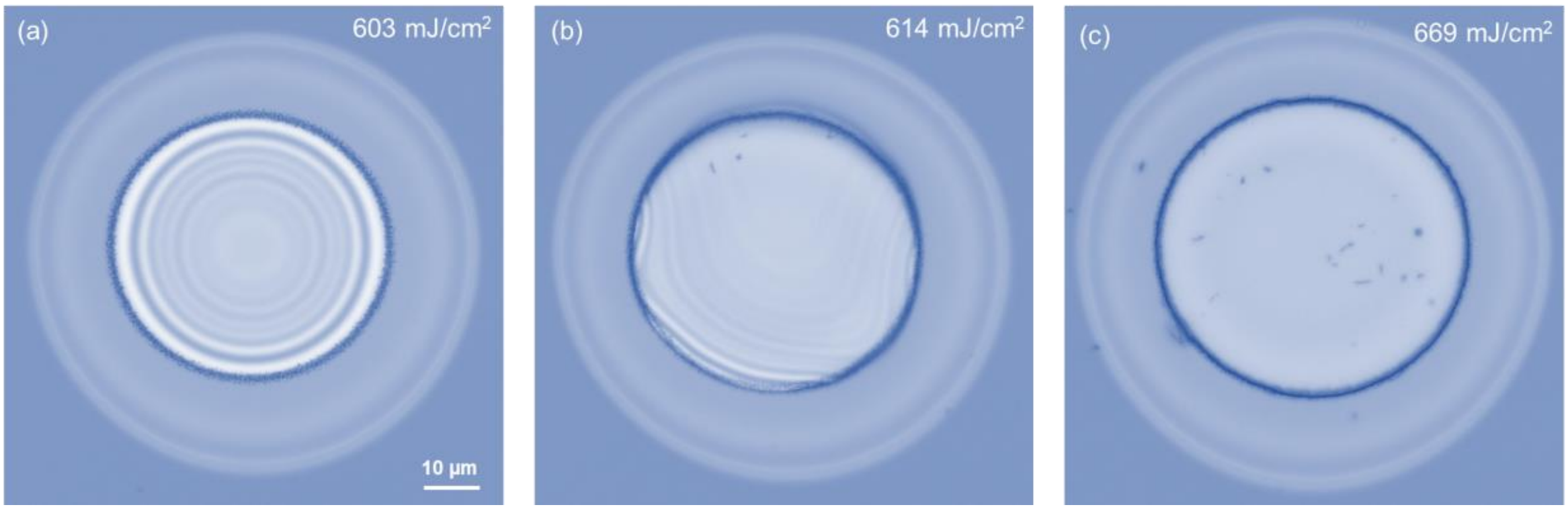

*Figure S5: Optical microscopy images of the microcavities cavity using $\lambda_{LED1}$ = 460 nm as illumination wavelengths, fabricated at laser fluences (a) below the damage threshold, (b) at a fluence generating partial fracture of the $SiO_2$ membrane and (c) a fluence triggering complete fracture.*

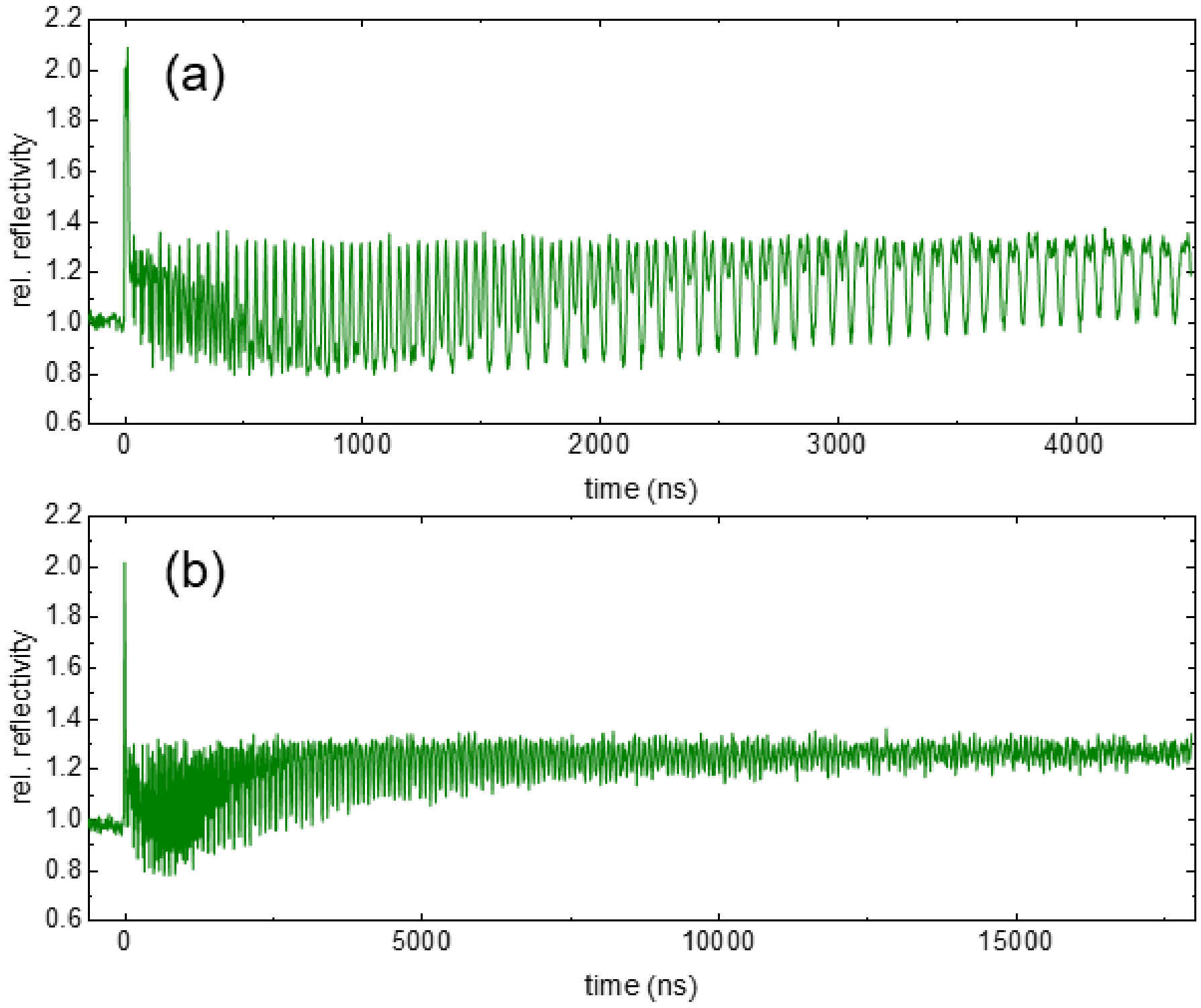

*Figure S6: Reflectivity evolution of the sample surface upon irradiation at F = 472 mJ/cm$^2$, leading to the formation of a microcavity. (a) shows the oscillation in a temporal window in which they can still be resolved temporally by the oscilloscope. (b) shows a larger window, in which the oscillation damping can be appreciated as well as oscillation lifetimes > 15 $\mu$s.*